# A deep machine learning potential for atomistic simulation of Fe-Si-O systems under Earth's outer core conditions

Chao Zhang,[1] Ling Tang,[2] Yang Sun,[3,4] Kai-Ming Ho,[3] Renata M. Wentzcovitch,[4] and Cai-Zhuang Wang[3*]

[1] Department of Physics, Yantai University, Yantai 264005, China

[2] Department of Applied Physics, College of Science, Zhejiang University of Technology, Hangzhou, 310023, China

[3] Department of Physics and Astronomy, Iowa State University, Ames, Iowa 50011, United States

[4] Department of Applied Physics and Applied Mathematics, Columbia University, New York, New York, 10027, United States

**Abstract**

Using artificial neural-network machine learning (ANN-ML) to generate interatomic potentials has been demonstrated to be a promising approach to address the long-standing challenge of accuracy versus efficiency in molecular dynamics (MD) simulations. Here, taking the Fe-Si-O system as a prototype, we show that accurate and transferable ANN-ML potentials can be developed for reliable MD simulations of materials at high-pressure and high-temperature conditions of the Earth's outer core. The ANN-ML potential for Fe-Si-O system is trained by fitting to the energies and forces of related binaries and ternary liquid structures at high pressures and temperatures obtained by first-principles calculations based on density functional theory (DFT). We show that the generated ANN-ML potential describes well the structure and dynamics of liquid phases of this complex system. The efficient ANN-ML potential with DFT accuracy provides a promising scheme for accurate atomistic simulations of structures and dynamics of complex Fe-Si-O system in the Earth's outer core.

## 1. Introduction

Molecular dynamics (MD) simulation has been demonstrated to be a very useful computational tool for investigating the structure and dynamics at an atomistic level of details for many systems in condensed matter physics, materials science, chemical and biological science, as well as earth science [1, 2]. However, to perform reliable MD simulations, accurate and efficient descriptions of interatomic forces are critical.

Quantum mechanics calculations based on first-principles density functional theory (DFT) can provide an accurate description of interatomic forces and total energies for many materials, and *ab initio* MD (AIMD) simulations based on DFT have been successful in studying the structures and dynamics of many materials [3, 4]. However, due to the heavy computational workload, AIMD can usually be performed with a small simulation cell size (usually less than 500 atoms) and shorter time (typically less than 1 ns) even with advanced supercomputers.

To overcome time length and system size limitations in MD simulations, considerable efforts in the past several decades have been devoted to developing empirical interatomic potentials for MD simulations of various classes of materials. Conventionally, such interatomic potentials are modeled by given mathematical functions with respect to atomic coordinates in the systems and contain some empirical parameters which to be fitted to the data from experimental measurement or first-principles calculations. Prototype interatomic potentials include Lennard-Jones potentials for noble gas and colloidal systems [5, 6], Tersoff and Stillinger-Weber potentials [7, 8] for covalent systems such as silicon and carbon, and embedded-atom method (EAM) potentials [9] for the metallic systems. Although these potentials have been widely used in MD simulations and have produced many useful results for better understanding the structures and properties of materials [10, 11], limitations for their application in more complex systems and/or under extreme environments have also been noticed. In many cases, reliable MD simulations for such complex systems are highly desirable when direct experimental studies become very difficult. For example, the Earth's outer core of our earth is believed to be composed of a liquid iron alloy with up to 10% of light elements such as silicon, oxygen, sulfur, carbon, or hydrogen. Despite extensive studies, chemical compositions and structures of the Earth's outer core are still elusive. Owing to the great pressures (135 – 363 GPa) and temperatures (3800 – 6500 K), experimental studies at core conditions are also limited. While MD simulations would provide useful insights into these problems, it is of great challenge to model interatomic potentials for such complex systems to ensure reliable MD simulations.

We note that it would be very difficult to choose appropriate mathematical functions for interatomic potentials based on chemical and physical intuition to correctly and efficiently describe the complicated interactions in complex materials [12]. On the other hand, machine learning (ML) is well-known for its ability in performing pattern recognition [13]. Since interatomic forces that govern atomic dynamics in a condensed matter system are predominately dependent on the positions and nature of neighboring atoms, interatomic potential fitting can be regarded as a pattern recognition problem and well suited for ML without assuming any mathematical functions. Within this spirit, considerable efforts in last several years have been devoted to the development of ML interatomic potentials for MD simulations of various materials [14-34]. Among various ML interatomic potentials schemes, deep learning based artificial neural network (ANN) first proposed by Behler and Parrinello [14] and further improved by Zhang *et al.* [30-33] has demonstrated to be very robust for reliable MD simulations of structures and behaviors many complex materials [35-41].

In this paper, we develop an ANN-ML interatomic potential for Fe-Si-O system, aimed at enabling accurate MD simulation of materials containing these three elements at extreme conditions of high pressure and high temperature similar to that in the Earth's outer core. We show that the developed ANN-ML interatomic potential describes well the structure and dynamics of the Fe-Si-O system at high pressures (> 100 GPa) and high temperature (> 3000 K). The potential will enable accurate and efficient atomistic simulations of structures and dynamics of complex Fe-Si-O systems in the Earth's outer core with a large number of atoms and longer simulation time.

The paper is organized as follows. In section 2, we describe the datasets and the detailed process and parameters used in the ANN-ML training. The training and testing accuracies in comparison with the first-principles DFT results are also discussed. Application of the developed Fe-Si-O ANN-ML potential to MD simulation studies of the structures and dynamics of Fe-Si and Fe-O binaries and Fe-Si-O ternaries at high-temperature liquid phases are presented in section 3 and section 4 respectively. Comparisons with available *ab initio* MD simulations are also discussed. Finally, a brief summary is given in section 5.

## 2. Development of ANN-ML potential for Fe-Si-O system

The data set used to train the ANN-ML interatomic potential for Fe-Si-O ternary system consists of high temperature and high pressure liquids of pure Fe and related binaries and ternary as listed in Table 1. These data are generated by AIMD simulations and are consists of potential energies for each structure and forces on every atom in the structures. The AIMD simulations are performed using Vienna Ab-Initio Simulation Package (VASP) [4, 42]. Projected-augmented-waves (PAW) with the Perdew-Burke-Ernzerhof (PBE) form of exchange-correlation potentials are adopted [43, 44]. Only the $\Gamma$ point is utilized to sample the Brillouin zone and the default energy cutoffs of 400 eV are employed. The AIMD simulations are carried out using the NVT ensemble with Nóse-Hoover thermostat under periodic boundary conditions. The time step of the AIMD simulations is 3 fs. A total 124455 snapshot structures with several different compositions from the AIMD have been collected for ML training, as can be seen from Table 1. The AIMD simulations for each Fe-Si-O system to collect the snapshots for ML are performed at 3800 K, 4000 K, 4300 K, and 4800 K respectively. At each temperature, the AIMD simulations for each system are performed at least with two different densities in the range specified in the Table 1.

**Table 1.** The training data sets used for the Fe-Si-O ANN-ML potential development. The RMSE of energy and force predicted by the ANN-ML model are the validation RMSE.

| System | Total number of atom | Total number of snapshot | density (g/cm$^3$) | Energy RMSE (meV/atom) | Force RMSE (meV/Å) |
|---|---|---|---|---|---|
| $Fe_{189}Si_{38}O_{23}$ | 250 | 23143 | 8.36 ~ 10.65 | 5.3 | 0.43 |
| $Fe_{189}Si_{61}$ | 250 | 29967 | 9.63 ~ 9.93 | 4.7 | 0.39 |
| $Fe_{189}O_{61}$ | 250 | 31906 | 8.79 ~ 10.49 | 6.0 | 0.47 |
| Si$_{80}$O$_{160}$ | 240 | 19283 | 4.98 ~ 6.19 | 5.4 | 0.31 |
| Fe | 256 | 20156 | 10.26 ~ 11.28 | 4.5 | 0.42 |

The DeepPot-SE model in the DeePMD-kit package [30] is applied in the training process to develop the ANN-ML potential for Fe-Si-O system. The cutoff radius of the model is set to 6.5 Å and descriptors decay smoothly from 6.0 Å to the cutoff radius of 6.5 Å. The size of the filter and fitting networks are (60, 120) and (240, 240, 240), respectively. A skip connection is built (ResNet) between two neighboring fitting layers [45]. The model is trained by the Adam stochastic gradient decent method [46] and the learning rate decreases exponentially with respect to the starting value of 0.001. The ANN is initialized with random numbers and the total number of training steps is 3,000,000. The decay rate and decay step are set to 0.96 and 10000, respectively. In addition, the prefactors in the loss functions are $p_e^{start} = 0.1$, $p_e^{limit} = 0.1$, $p_f^{start} = 1000$, $p_f^{limit} = 1$, $p_v^{start} = 0$, $p_v^{limit} = 0$.

Fig. 1 shows the comparison of total potential energies and forces on each atom from the trained ANN-ML potential and *ab initio* calculated results for Fe$_{189}$Si$_{38}$O$_{23}$ liquid. The energies and forces predicted by the ANN-ML model and calculated by *ab initio* method are plotted in the same figure as vertical coordinate and horizontal coordinate, respectively. The root mean square error (RMSE) of energy is about 5.3 meV/atom and the force RMSE is about 0.43 eV/Å for the validation data. The training accuracy for other systems listed in Table 1 is similar to the one shown in Fig. 1. Hence, the trained ANN-ML potential has an accuracy comparable to *ab initio* calculations.

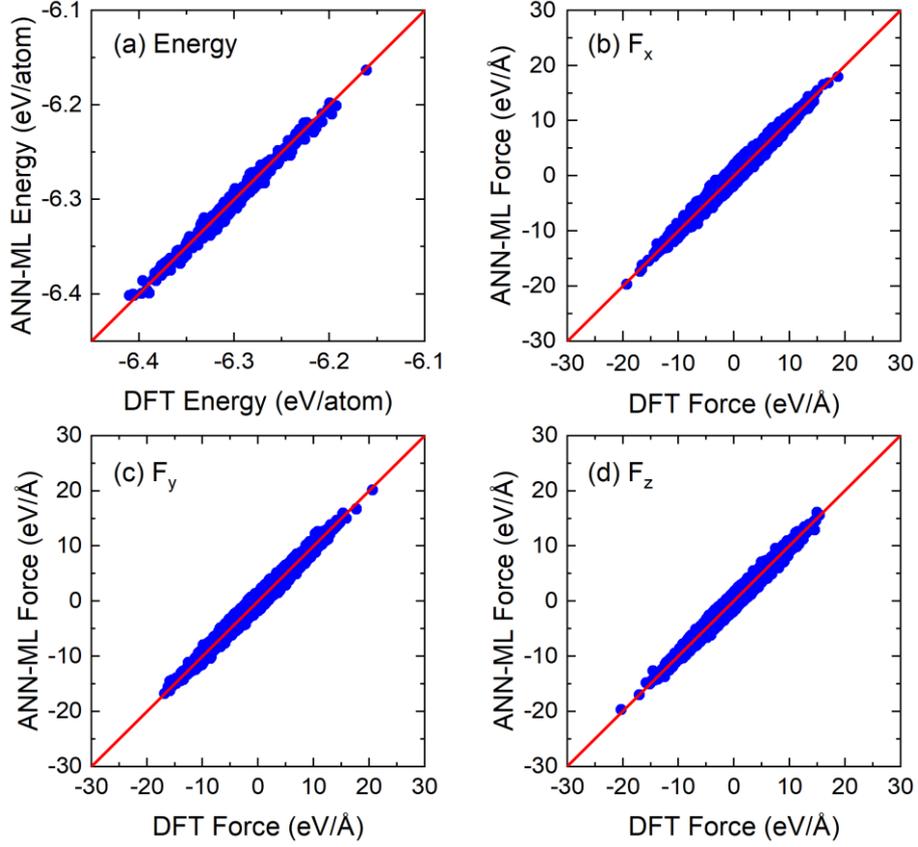

**Fig. 1** ANN-ML versus DFT energies and forces for the validation data set of $Fe_{189}Si_{38}O_{23}$.

## 3. MD simulation of Fe-Si, Fe-O, and Si-O binary liquids

With the interface of the DeePMD-kit to the LAMMPS code [47], MD simulations can be directly performed with the generated ANN-ML potential [30]. We first validate the reliability of the developed ANN-ML potential by compare the structures and dynamics of $Fe_{189}Si_{61}$, $Fe_{189}O_{61}$, and $Si_{80}O_{160}$ liquids obtained from the MD simulations using the developed ANN-ML potential with those from AIMD simulations. The MD simulations by the ANN-ML potential are performed using a NVT ensemble and a Nóse-Hoover thermostat. Small and large simulation cells are used in the ANN-ML potential MD simulations. For the small simulation cell, the same box length as in the AIMD simulation is used. The large cell is a 2 × 2 × 2 supercell of the small one, thus the density of the small and large cells are same. Periodic boundary conditions are applied in the three directions and the time step of the simulations is 3 fs. According to the size of simulation cell, we refer to small simulation cell of 200 ~ 256 atoms as "S" and large simulation cell of 2000 ~ 5000 atoms as "L". For example, AIMD$_S$ model of Fe$_{189}$Si$_{38}$O$_{23}$ system contains 240 atoms, whereas ANN-MD$_L$ model of Fe$_{189}$Si$_{38}$O$_{23}$ system contains 2000 atoms. The same simulation conditions are applied, i.e., initial configurations, simulation steps, and NVT ensemble, on AIMD$_S$ and ANN-MD$_S$ models. The density of the studied Fe$_{189}$Si$_{61}$, Fe$_{189}$O$_{61}$, and Si$_{80}$O$_{160}$ liquid in this section are 9.78 g/cm$^3$, 9.57 g/cm$^3$, and 5.31 g/cm$^3$ at 3800 K, respectively. The pressures obtained from the ANN-MD$_S$ and ANN-MD$_L$ models are almost same, but are about 6.8% ~ 10.1% larger than that from the AIMD$_S$ model. The pressures of the Fe$_{189}$Si$_{61}$, Fe$_{189}$O$_{61}$, and Si$_{80}$O$_{160}$ liquid are 135 GPa (144 GPa), 134 GPa (147 GPa), and 133 GPa (143 GPa) from AIMD$_S$ (ANN-MD$_S$) model at 3800 K, respectively.

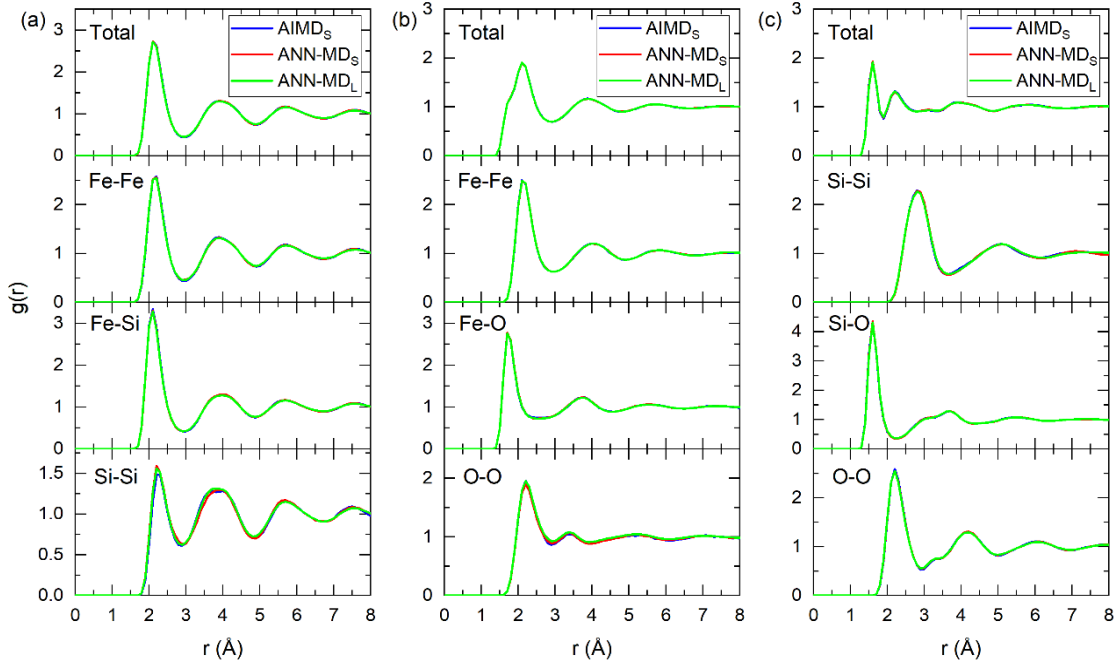

**Fig. 2** Total and partial pair correlation functions of liquid (a) $Fe_{189}Si_{61}$, (b) $Fe_{189}O_{61}$, and (c) $Si_{80}O_{160}$ at 3800K.

The structures of the $Fe_{189}Si_{61}$, $Fe_{189}O_{61}$, and $Si_{80}O_{160}$ liquid at 3800 K are analyzed from the $AIMD_S$, $ANN-MD_S$, and $ANN-MD_L$ model. Total and partial pair correlation functions (PCFs) are shown in Fig. 2. The PCFs of binary $Fe_{189}Si_{61}$, $Fe_{189}O_{61}$, and $Si_{80}O_{160}$ liquids at 3800 K obtained by AIMD and ANN-MD agree well with each other. For $Fe_{189}Si_{61}$ binary system, the position of the first peak of partial PCFs of Fe-Fe, Fe-Si, and Si-Si are approximately 2.2 Å, 2.1 Å, and 2.2 Å, respectively. These results indicate that the nearest neighbor distances between Fe and Si atoms and among the Fe or Si atoms themselves in $Fe_{189}Si_{61}$ binary liquid are very similar, which is a good agreement with previous works [48, 49]. For $Fe_{189}O_{61}$ binary system, the first PCF peak of O-O is 2.2 Å, which is significantly larger than the Fe-O (1.7 Å) and Fe-Fe (2.1 Å). This indicates that O atoms do not form the nearest neighbor bonds among themselves in liquid Fe-O system. The first PCF peak of O-O in $Si_{80}O_{160}$ binary system is also 2.2 Å, which stands the medium of Si-Si (2.8 Å) and Si-O (1.6 Å). It is noteworthy that the bond length of O-O in $Si_{80}O_{160}$ system is same to that in $Fe_{189}O_{61}$ system, indicating that O atoms also do not form the nearest neighbor bonds among themselves in the liquid Si-O system. Besides the PCF, the partial angular distribution functions (ADFs) can provide more local structural information about the liquid samples. The ADFs obtained from AIMD and ANN-MD coincide with each other, as shown in the Supporting Information, which further indicates the reliability of ANN-ML potential. Excellent agreement of PCFs and ADFs is also observed for the AIMD with small (ANN-MD$_S$) and large (ANN-MD$_L$) simulations cells, therefore confirming the validation of ANN-ML potential for large systems.

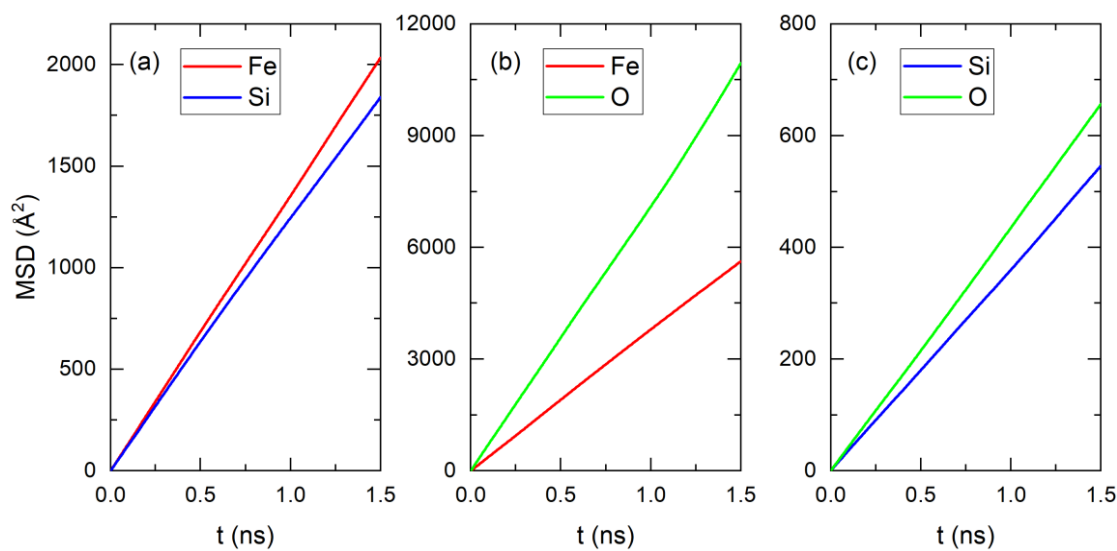

**Fig. 3** Mean square displacement of (a) $Fe_{189}Si_{61}$, (b) $Fe_{189}O_{61}$, and (c) $Si_{80}O_{160}$ liquid from ANN-MD$_L$ model at 3800 K.

To quantitatively study the dynamic properties, we calculated the self-diffusion constants $D$ of every elements in the binary liquids. The mean-square displacement (MSD) as a function of time is given by [50, 51]

$$\langle R_\alpha^2(t)\rangle = \frac{1}{N_\alpha}\langle\sum_{i=1}^{N_\alpha}|R_{i\alpha}(t+\tau) - R_{i\alpha}(\tau)|^2\rangle,$$

where $N_\alpha$ is the total atomic number of α species, $R_{i\alpha}$ is the coordinates of the atom $i$, and $\tau$ is the arbitrary origin of time. The MSD of the liquids in the limit of long time should behave linearly with the time, and the slope of the line gives the self-diffusion constant $D$ by the Einstein relationship,

$$D = \lim_{t\to\infty}\langle R_{i\alpha}^2(t)\rangle/6t.$$

The self-diffusion constant $D$ of $Fe_{189}Si_{61}$, $Fe_{189}O_{61}$, and $Si_{80}O_{160}$ binary systems are calculated within 1.5 ns from the ANN-MD$_L$ models. For $Fe_{189}Si_{61}$ binary system, Fe and Si have similar diffusing constant, i.e., $D_{Fe} = 0.26 \times 10^{-8}$ m$^2$/s, and $D_{Si} = 0.20 \times 10^{-8}$ m$^2$/s. The diffusion constants of Fe and O in $Fe_{189}O_{61}$ binary system is $D_{Fe} = 0.63 \times 10^{-8}$ m$^2$/s and $D_O = 1.20 \times 10^{-8}$ m$^2$/s, respectively, which means O atoms move faster than Fe atoms in $Fe_{189}O_{61}$ system. The diffusion constants of Si and O in $Si_{80}O_{160}$ binary system is $D_{Si} = 0.60 \times 10^{-8}$ m$^2$/s and $D_O = 0.73 \times 10^{-8}$ m$^2$/s, respectively. These data agree with those obtained from liquid Fe and Fe-O under Earth's outer core conditions [49, 52-56].

## 4. MD simulation of Fe-Si-O ternary liquids

### 4.1 Validation of ANN-ML potential for $Fe_{189}Si_{38}O_{23}$ ternary liquid

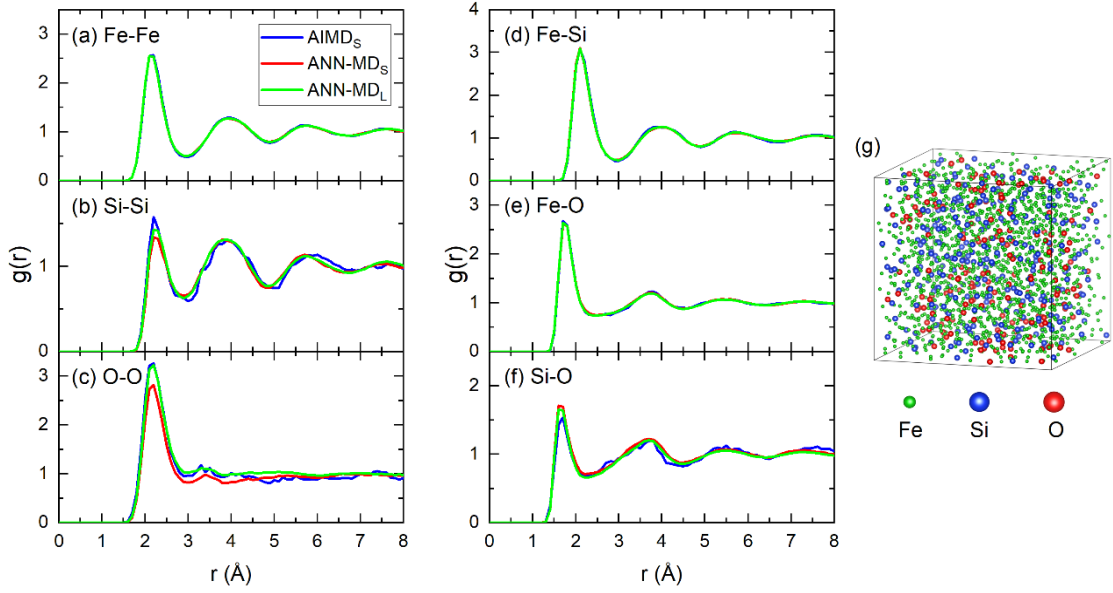

**Fig. 4** (a)-(f) Partial pair correlation functions of liquid $Fe_{189}Si_{38}O_{23}$ at 3800 K. (g) Snapshot of $Fe_{189}Si_{38}O_{23}$ liquid from ANN-$MD_L$ model at 3800 K and 145 GPa at 3.0 ns.

We performed MD simulation of $Fe_{189}Si_{38}O_{23}$ ternary system with density of 9.91 g/cm$^3$ at 3800 K. The pressures of the $Fe_{189}Si_{38}O_{23}$ liquid at 3800 K are 132 GPa, 145 GPa, and 145 GPa from AIMD$_S$, ANN-MD$_S$, and ANN-MD$_L$ model, respectively. The partial PCFs of $Fe_{189}Si_{38}O_{23}$ ternary system at 3800 K from AIMD$_S$, ANN-MD$_S$, and ANN-MD$_L$ models are shown in Fig. 4. The partial PCFs distributions from AIMD$_S$, ANN-MD$_S$, and ANN-MD$_L$ models are similar, especially the peak positions. This further indicates the validation of ANN-ML potential for ternary systems. The first peak of O-O partial PCF is located at 2.2 Å, whereas the positions of the first peak of Fe-O and Si-O are both 1.7 Å. This indicates O atoms do not form nearest-neighbor bonds among themselves in the Fe-Si-O ternary system. The bond lengths of O-O, Fe-O, and Si-O in the $Fe_{189}Si_{38}O_{23}$ ternary system are similar to those in $Fe_{189}O_{61}$ and $Si_{80}O_{160}$ binary systems. The bond lengths of Fe-Fe, Si-Si, and Fe-Si in the $Fe_{189}Si_{38}O_{23}$ ternary system are 2.2 Å, 2.2 Å, and 2.1 Å, respectively, which are similar to that in the $Fe_{189}Si_{61}$ binary system. In addition, the potential energy fluctuates around a constant value with MD simulation time, indicating no phase transition or separation takes place at 3800 K, as shown in the Supporting Information. Fig. 4(g) shows that Fe, Si, and O atoms are mixed well with each other. This indicates there is no phase separation in Fe-Si-O system and no immiscibility between Fe-Si and Fe-Si-O liquids at 3800 K, which is consistent with previous work [49].

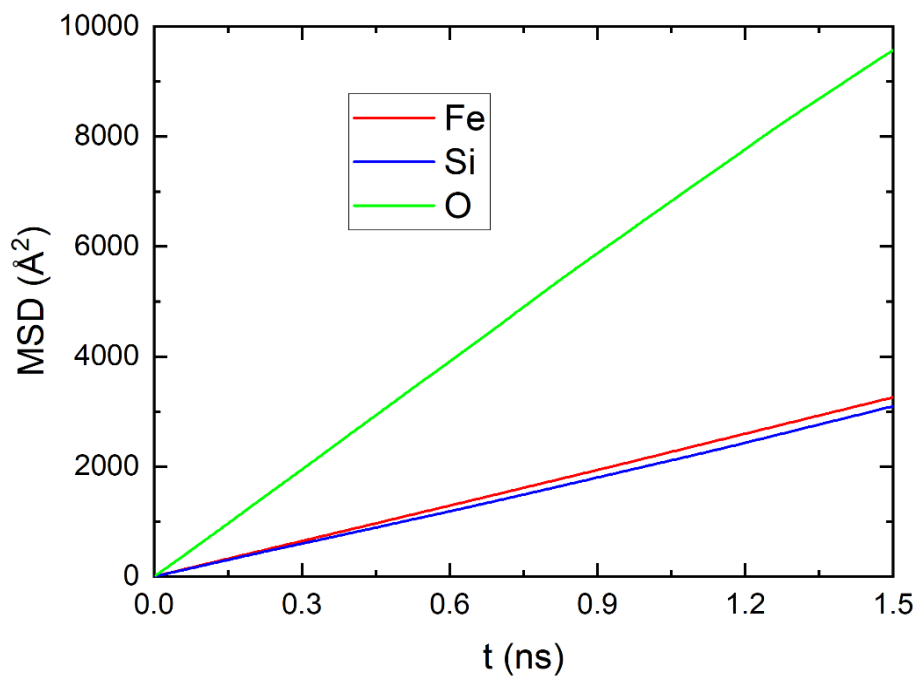

**Fig. 5** Mean square displacement of liquid $Fe_{189}Si_{38}O_{23}$ at 3800 K from ANN-MD$_L$ model.

The self-diffusion constant $D$ of $Fe_{189}Si_{38}O_{23}$ ternary system at 3800 K are calculated within 1.5 ns from the ANN-MD$_L$ model. We find $D_{Fe} = 0.36 \times 10^{-8}$ m$^2$/s, $D_{Si} = 0.34 \times$ $10^{-8}$ m$^2$/s, and $D_O = 1.07 \times 10^{-8}$ m$^2$/s. The self-diffusion constants of Fe and Si in the Fe$_{189}$Si$_{38}$O$_{23}$ ternary system are similar, which is also found in Fe$_{189}$O$_{61}$ binary system. $D_O$ is about three times of D$_{Fe}$ or D$_{Si}$, which is consistent with previous work [49].

## 4.2 Further test of ANN-ML potential for Fe$_{189}$Si$_{23}$O$_{38}$ and Fe$_{158}$Si$_{14}$O$_{28}$ ternary liquids

In addition to the Fe$_{189}$Si$_{38}$O$_{23}$ ternary system at 3800 K whose AIMD snapshots have been used in the training data for the ANN-ML potential, we also performed MD simulations for liquids Fe$_{189}$Si$_{23}$O$_{38}$ and Fe$_{158}$Si$_{14}$O$_{28}$ ternary systems (whose AIMD snapshots are not used in the ANN-ML training) to further test the accuracy and transferability of the obtained ANN-ML potential for Fe-Si-O systems with O richer than Si.

The density of Fe$_{189}$Si$_{23}$O$_{38}$ liquid at 4800 K is 9.85 g/cm$^3$ for the AIMD$_S$, ANN-MD$_S$, and ANN-MD$_L$ models, whereas the pressures are 132 GPa, 143GPa, and 143 GPa from the three models, respectively. The partial PCFs of the Fe$_{189}$Si$_{23}$O$_{38}$ from ANN-MD$_S$ model (250 atoms) agrees well with that from AIMD$_S$ model (250 atoms), as shown in Fig. 6. From Fig. 6(g), we can see that Fe, Si, and O atoms mix well with each, which means no phase separation in Fe$_{189}$Si$_{23}$O$_{38}$ ternary system.

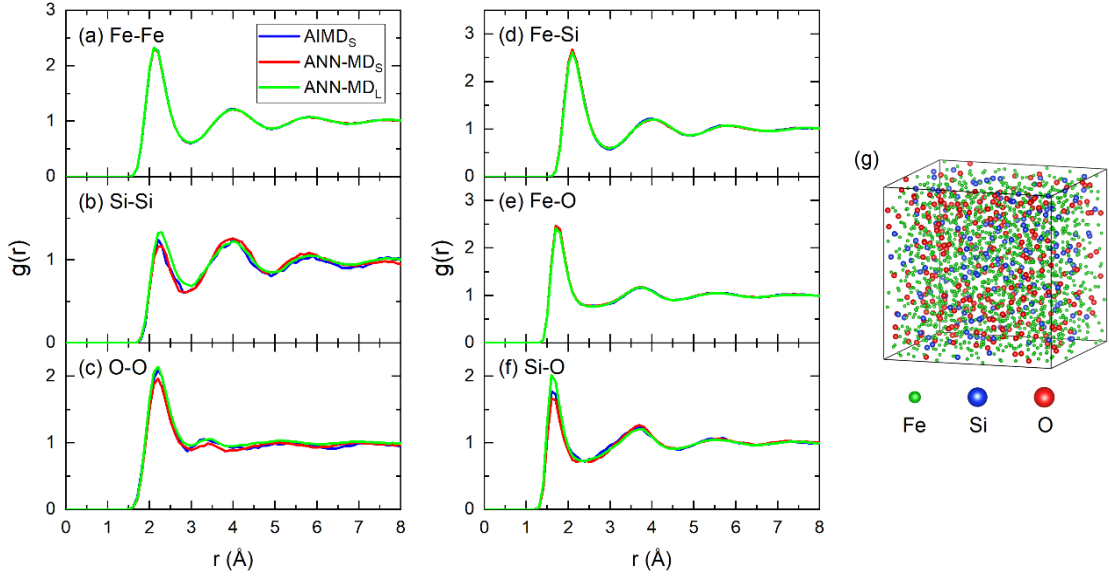

**Fig. 6** (a)-(f) Partial pair correlation functions of $Fe_{189}Si_{23}O_{38}$ liquid at 4800 K. (g) Snapshot of $Fe_{189}Si_{38}O_{23}$ liquid from ANN-MD$_L$ at 4800 K and 145 GPa at 3.0 ns.

The density of $Fe_{158}Si_{14}O_{28}$ liquid at 4500 K is 10.16 g/cm$^3$ for the AIMD$_S$, ANN-MD$_S$, and ANN-MD$_L$ models, whereas the pressures are 136 GPa, 148 GPa, and 148 GPa from the three models, respectively. The partial PCFs of the $Fe_{158}Si_{14}O_{28}$ from AIMD$_S$ model (200 atoms) agrees well with that from ANN-MD$_S$ and ANN-MD$_L$ models (5000 atoms), as shown in Fig. 7. The efficiency of the ANN-ML potential enables MD simulations with larger unit cell to compare with the results from the small (200 atoms) unit cell. From Fig. 7(g), we can see that Fe, Si, and O atoms mix well with each, which means no phase separation in $Fe_{158}Si_{14}O_{28}$ ternary system.

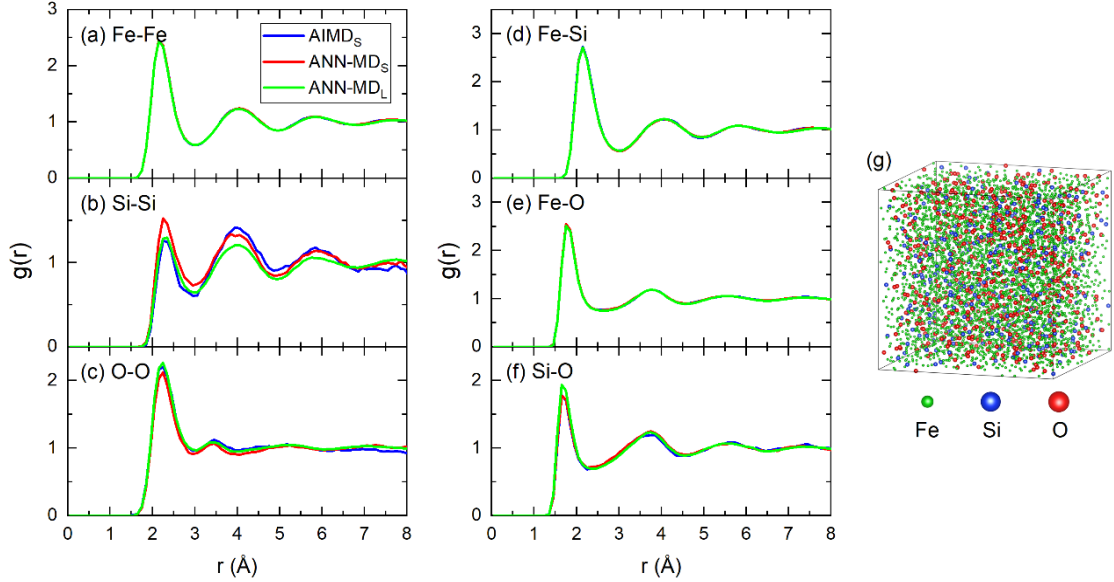

**Fig.7** (a)-(f) Partial pair correlation functions of $Fe_{158}Si_{14}O_{28}$ liquid at 4500 K. (g) Snapshot of $Fe_{158}Si_{14}O_{28}$ liquid from ANN-MD$_L$ at 4500 K and 148 GPa at 20 ps.

## 5. Summary

In this paper, we have developed an ANN-ML potential for Fe-Si-O system at the high-pressure and high-temperature conditions of the Earth's outer core using the DeePMD-kit and VASP software packages. The developed ANN-ML potential can be used in the LAMMPs package to perform MD simulations. The ANN-ML potential not only can well reproduce the AIMD results on structures of the binary and ternary liquids whose snapshot structures were included in the ANN-ML train data set, but also provide consistent MD simulation results of ternary liquids ($Fe_{189}Si_{23}O_{38}$ and $Fe_{158}Si_{14}O_{28}$) whose snapshot structures were not included in the train data set. The results show there is no phase separation and exsolution in our studied three ternary system ($Fe_{189}Si_{38}O_{23}$, $Fe_{189}Si_{23}O_{38}$, and $Fe_{158}Si_{14}O_{28}$) around 136 GPa. Our results suggest the ANN-ML potential would be a promising avenue for MD simulation of complex Fe-Si-O systems under the Earth's outer core conditions.

**Acknowledgements**

C. Zhang was supported by the National Natural Science Foundation of China (Grants Nos. 11874318 and 11774299) and the Natural Science Foundation of Shandong Province (Grants Nos. ZR2018MA043 and ZR2017MA033). L. Tang acknowledges the support by the National Natural Science Foundation of China (Grant No. 11304279). Y. Sun was supported by US NSF Awards No. EAR-1918134 and No. EAR-1918126. K. M. Ho and C. Z. Wang were supported by US NSF Awards No. EAR-1918134. R. M. Wentzcovitch was supported by US NSF Awards No. EAR-1918126. .